# Microkernel-Based Web Architecture: Design & Implementation Considerations

Vick Dini
*Department of Electronics, Information, and Bioengineering*
*Politecnico di Milano*
Milan, Italy
vickpierce.dini@polimi.it

*Abstract*—In this vision paper I propose a middle-ground alternative between monolithic and microservice web architectures. After identifying the key challenges associated with microservice architectures, I revised the design of a microkernel-based web architecture, considering these challenges as well as recent architectural advancements. Next, I examined contemporary approaches to various self-* properties and explored how this new architecture could enhance them, including a modified version of the MAPE-K loop. Once the high-level design of the microkernel architecture was finalized, I evaluated its potential to address the identified challenges. Lastly, I reflected on several implementation aspects of the proposed work.

## I. INTRODUCTION

Microservices evolved from earlier paradigms like the Service-Oriented Architecture (SOA) to address issues in agility, scalability, and deployment, especially with the rise of cloud computing [1]. By the 2010s, many organizations began transitioning from monolithic applications to microservices, despite the technical challenges involved, with promises of improved maintainability and flexibility [2]. However, microservices still face challenges related to increased system complexity, design, and service integration, sparking ongoing research into solutions like container orchestration and service optimization [3].

In the meantime, [4] worked on a microkernel-based web architecture to tackle the challenges of web monoliths. Although it didn't become as popular as microservices, an updated version might be useful to solve today's problems.

In fact, in this paper I identify the main challenges of microservice architectures, reconsider the microkernel-based web architecture proposed in [4] and analyze how it could be updated to help overcome such challenges.

In the next sections, I pose some research questions (Section II) and provide answers by briefly reviewing the literature (Section III), analyzing the architecture proposed in [4], updating it to meet today's needs (Section IV), evaluating several of its self-* properties, reflecting on how this approach tackles the main challenges of microservices, and making some considerations on the implementation of this novel architecture, after which I conclude this paper (Section VIII).

## II. RESEARCH QUESTIONS

Throughout this research I focused on the following questions:

- RQ1: What are the key challenges involved in adopting, implementing, and maintaining a microservices architecture?
- RQ2: How could a microkernel-based web architecture help overcome such challenges?

I addressed RQ1 by conducting a brief literature review to capture the most recent advancements in the field. Subsequently, I provide an answer for RQ2 by drawing on these findings and building on the architecture proposed by [4].

## III. MICROSERVICES

The following challenges involved in adopting, implementing, and maintaining microservice architectures emerged from the literature review:

1) **Security Risks**: Microservices introduce security challenges due to their distributed nature and the complex communication between services. Compromising one microservice can expose the entire application to threats [5] [6].
2) **Performance Overhead**: Microservices can incur performance penalties due to the overhead of managing many independent services and their intercommunication [7].
3) **Service Deployment and Monitoring**: Managing deployment and monitoring of numerous microservices in different environments is a complex task, necessitating robust automation tools [8].
4) **Service Interaction and Communication**: Understanding and visualizing the interaction between multiple microservices is a significant challenge, especially as the architecture scales [9].
5) **Finding the Right Granularity**: Determining the appropriate size and scope of microservices is difficult, and poor decisions on service granularity can lead to inefficiency and increased complexity [10] [11].
6) **Data Management**: Managing data consistency across microservices is difficult due to distributed data stores, which can result in issues like eventual consistency or complex transaction management [12].

7) **Troubleshooting**: Debugging microservices presents unique challenges due to their distributed and dynamic nature, including their distributed nature [13], complex service communication [14], state reproduction [15], trace and visualization tools [16], and fault localization [17].

## IV. Microkernel-Based Web Architecture

Given that *microkernel* ($\mu$K) is the central concept of this paper, I synthesized the definitions provided by [18] and [19] into the following unified definition:

> The microkernel approach reduces the kernel to only the basic functions needed for system operation, like managing address spaces, interprocess communication (IPC), and scheduling. Other tasks, including device drivers, are handled by separate servers that run in user mode and operate in their own address spaces. This setup helps protect these servers from each other. The microkernel focuses on providing essential services, while additional functions are managed by independent kernel and system server processes.

Based on this definition, I propose the microkernel-based web architecture in Fig. 1. From a certain perspective, it may be thought of as a web operating system running on top of the traditional operating system. As shown, the $\mu$K and adjacent components would perform the most system-related activities (Fig. 2), such as managing the interactions with system resources and general purpose utilities, whereas the modules would be in charge of the application logic.

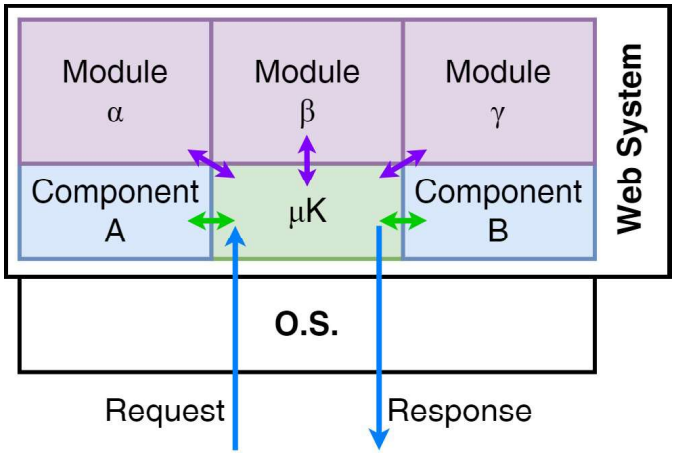

Fig. 1. Microkernel-Based Web Architecture

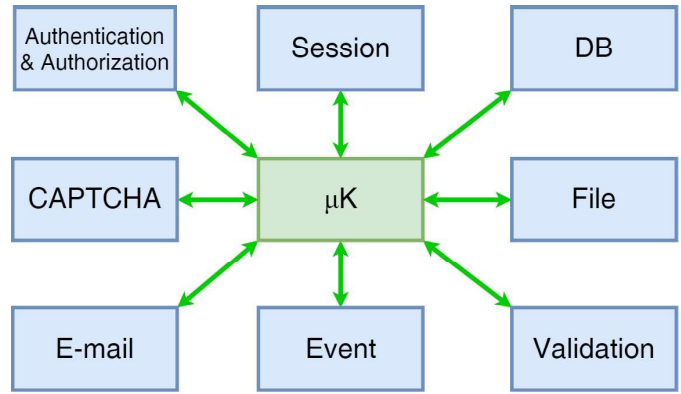

Fig. 2. $\mu$K and Components

### A. Components

The components proposed in Fig. 2 could be described as follows:

- **Authentication & Authorization**: The authentication and authorization component verifies user identities through methods such as username/password, multi-factor authentication (MFA), or OAuth, while also implementing role-based access control (RBAC) to assign roles and define permissions for resource access. It fine-tunes access control with specific permissions for actions like reading, writing, or deleting resources, and generates secure tokens for authenticated users to facilitate stateless communication between client and server. Additionally, it maintains audit logs of authentication attempts, access events, and changes to user roles or permissions for security audits and compliance, provides mechanisms for password resets and account recovery to ensure security, and allows single sign-on (SSO) for users to access multiple applications with a single authentication. It also implements rate limiting to protect against brute-force attacks by restricting login attempts.
- **CAPTCHA**: A CAPTCHA component helps distinguish between human users and automated bots by presenting challenges that are easy for humans but difficult for machines to solve. It creates various types of challenges, such as image recognition, text-based puzzles, or interactive tasks, and evaluates user responses to determine whether access should be granted or denied. The component can adjust the complexity of challenges based on user behavior, session history, or previous attempts to enhance user experience. It seamlessly integrates with user registration, login, and comment forms to prevent spam and unauthorized submissions, while also providing alternative challenge formats for users with disabilities to ensure compliance with accessibility standards. Additionally, it tracks and reports on CAPTCHA usage, including success rates and attempts, to improve effectiveness and user experience, and analyzes patterns of user behavior and IP addresses to flag suspicious activity that may require further verification.
- **E-mail**: The email component facilitates the sending of emails to users, including welcome messages, notifications, and transactional emails, while supporting both HTML and plain text formats. It allows for the creation and management of email templates to ensure consistent branding and messaging, and enables dynamic content insertion to personalize emails based on user data, such as names and preferences. The component monitors email delivery, open rates, and click-through rates to provide insights into user engagement and campaign effectiveness, and manages bounced emails by categorizing them as soft or hard bounces, automatically removing invalid addresses from future mailings. It ensures compliance with regulations like CAN-SPAM and GDPR by providing options for users to opt out of communications and clearly identifying the sender. Additionally, it implements a queuing system to handle large volumes of emails, offers robust error handling and logging for troubleshooting delivery issues, connects with other services for automated email communications based on user actions, and includes security measures like DKIM and SPF to prevent spoofing and ensure secure delivery.

- **Session**: The session component is essential for managing user sessions and maintaining state across interactions. It facilitates the creation of user sessions upon login, generating unique session identifiers to track user activity. The component manages session storage, allowing for the persistence of user data, preferences, and temporary states while the user is active. It handles session expiration and renewal, ensuring that sessions automatically expire after a specified period of inactivity to enhance security. Additionally, it provides mechanisms for session invalidation, enabling users to log out and terminate their sessions securely. The component may also support multi-device session management, allowing users to access their accounts from different devices while maintaining session continuity. Furthermore, it ensures compliance with security standards by implementing measures like HTTPS, session hijacking prevention, and secure cookie management.
- **Event**: The event component allows developers to define and create custom events that can be triggered by specific actions, such as user interactions, system changes, or scheduled tasks. It enables parts of the application to register event listeners that respond to these events, facilitating asynchronous communication between components. The component manages how events bubble up or are captured within the application, allowing for flexible handling at different levels of the component hierarchy. It implements a queue to manage events, ensuring they are processed in the order they are triggered, which is particularly useful in high-traffic situations. Additionally, it provides robust error handling mechanisms for issues that may occur during event processing, including logging and notifications. The component can integrate with other systems, such as notification or logging frameworks, to trigger actions in response to events, and supports techniques like debouncing and throttling to optimize event handling and improve user experience. It also allows for the scheduling of events to occur at specific times or intervals, tracks and reports on event occurrences to provide insights into user behavior and system performance, and manages the state related to events, enabling the application to respond appropriately based on current conditions or interactions.
- **Database (DB)**: The database component establishes and manages connections to the databases, handling connection pooling to optimize resource usage while facilitating the execution of database queries, including complex joins, aggregations, and transactions, with proper result handling. It implements validation rules to ensure data integrity and consistency before records are inserted or updated, and supports the creation and modification of database schemas to accommodate evolving data structures. The component manages indexes to improve query performance by optimizing data retrieval speeds and provides mechanisms for data backup and recovery to protect against data loss. Additionally, it facilitates seamless database migrations for schema changes and tracks database interactions through logging and monitoring to ensure efficient operation.
- **File**: The file component facilitates the uploading of files from users to the server, supporting multiple file uploads and progress tracking, while also enabling users to download files stored on the server with proper handling of different file types and formats. It allows for the creation of directories or folders to logically organize files, making navigation and management easier. The component manages metadata associated with files, such as size, type, upload date, and permissions, and implements access control mechanisms to restrict who can view, edit, or delete files based on user roles or attributes. Additionally, it supports versioning of files to track changes over time, allowing users to revert to previous versions if needed, and provides search functionality to help users quickly locate files based on names, types, or metadata. Robust error handling is included for file operations, such as failed uploads or downloads, to ensure a smooth user experience, and it offers mechanisms for backing up files and recovering them in case of accidental deletion or data loss.
- **Validation**: The validation component verifies that user inputs meet specified criteria, such as data types, formats, lengths, and required fields, to prevent invalid or harmful data from entering the system. It checks relationships between multiple fields to ensure data coherence, such as confirming that the end date is later than the start date, and provides clear, informative error messages when validation fails, helping users understand necessary corrections. The component performs server-side validation to ensure data integrity regardless of client-side manipulations, enhancing security, while also implementing client-side checks for immediate feedback before data submission to improve user experience. It allows developers to define custom validation rules for specific business logic requirements, ensuring flexibility, and cleans and sanitizes inputs to remove potentially harmful characters or scripts, helping prevent security vulnerabilities like SQL injection or cross-site scripting (XSS). Additionally, it supports asynchronous validation through AJAX requests to verify data against external resources, integrates with existing frameworks or libraries to streamline validation processes, and tracks and logs validation errors for monitoring purposes to identify common issues that may need addressing.

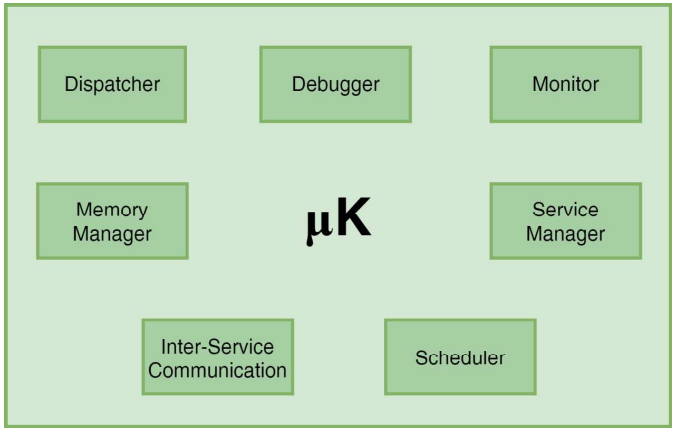

Fig. 3. $\mu$K

## B. Microkernel

As for the $\mu$K (Fig. 3), the most essential features it could comprise include:

- **Dispatcher**: When a user makes a request through a web browser, the dispatcher determines its destination and forwards it to the appropriate controller, module, or service responsible for processing. Additionally, the dispatcher serves as a load balancer by distributing incoming requests across multiple instances of web servers or application components. This dual function optimizes performance and enhances scalability, ensuring that no single server is overloaded while effectively directing the flow of requests.
- **Memory Manager**: The memory manager features robust virtual memory management, which abstracts memory to provide each service with its own address space, ensuring isolation and protection. It includes memory protection to prevent services from accessing each other's memory, safeguarding against accidental or malicious interference. Additionally, it offers dynamic memory allocation for on-the-fly allocation and freeing of memory, adapting to the needs of services, along with swapping capabilities to manage the movement of services in and out of physical memory, allowing for larger workloads than can fit in RAM. The manager also supports shared memory, enabling services to efficiently communicate through shared segments, and incorporates performance monitoring tools to track memory usage and optimize resource allocation.
- **Inter-Service Communication**: The Inter-Service Communication Manager features robust RESTful API support, enabling easy integration and communication between services. It facilitates real-time communication through WebSockets, allowing for bidirectional interaction between services and clients without the need for polling. The component integrates message queuing systems to decouple services and manage asynchronous message delivery efficiently. Load balancing techniques distribute requests among multiple service instances, enhancing scalability and reliability. Furthermore, the Inter-Service Communication Manager offers data serialization support for formats like JSON and XML to facilitate seamless data exchange, and implements robust error handling and automatic retry mechanisms to ensure reliable communication even in the face of transient failures.
- **Debugger**: The debugger offers real-time debugging, allowing developers to live-debug running services and modify code execution on the fly. It provides step-through execution to observe the flow of code line by line, and supports breakpoints to pause execution at specific lines or conditions, as well as watchpoints to monitor variable values. Integrated logging capabilities track service interactions and errors, aiding in issue diagnosis based on historical data. Remote debugging enables troubleshooting of services on remote servers, while variable inspection tools help identify incorrect states in real time. Developers can examine the call stack to understand the sequence of function calls and utilize memory inspection to detect leaks or allocation issues. The debugger also features automated error reporting and analysis, presenting error reports with stack traces for quick diagnosis. Performance profiling tools identify bottlenecks and analyze resource consumption, and API testing integration facilitates testing of service endpoints directly within the debugger. Additionally, a user-friendly interface allows developers to visualize service architectures and their interactions, making issue identification more straightforward.
- **Monitor**: The monitor features continuous service health monitoring to ensure services are running correctly, alerting administrators to any failures or performance issues. It collects and displays key performance metrics, such as response times, throughput, and resource utilization (CPU, memory, disk I/O) for each service, while sending real-time alerts for critical issues like downtimes, high resource usage, or errors to enable prompt responses. The monitor aggregates logs from various services, providing a comprehensive view of service interactions and errors for easier analysis, and offers a user-friendly dashboard to visualize overall health and performance through graphs and charts. Historical performance data is stored for trend analysis, helping administrators identify patterns and make informed decisions. Additionally, it monitors security events and anomalies to enhance the environment's security posture, tracks configuration changes to enable rollback capabilities, and visualizes service dependencies to aid in root cause analysis. Scalability management en-sures effective resource allocation with recommendations for adjusting services based on demand, while API usage tracking identifies potential bottlenecks or underutilized services. Finally, customizable alerts and reports allow administrators to set specific thresholds and generate tailored monitoring solutions.
- **Service Manager**: The service manager offers automatic service discovery, which detects and registers services within the environment to enable dynamic communication and integration. It facilitates service deployment, including configuration management, version control, and the ability to roll back to previous versions if necessary. The manager continuously monitors the status and performance of services through regular health checks, ensuring they function correctly, and provides tools for scaling services up or down based on demand to optimize resource allocation and performance. It centrally manages service configurations, allowing for easy updates and consistency across environments. Additionally, it collects logs from services and tracks changes for auditing purposes, while overseeing APIs to ensure secure and efficient interactions through rate limiting, authentication, and documentation. Dependency management visualizes service dependencies, aiding in understanding the impact of changes and maintaining stability, and automated recovery mechanisms minimize downtime by restoring failed services. Finally, the service manager offers a user-friendly web interface that provides dashboards, reports, and alerts to assist administrators in managing the overall system effectively.
- **Scheduler**: The scheduler is responsible for task prioritization by assigning priorities based on importance and urgency, ensuring critical tasks are executed first. It employs time slicing to fairly allocate CPU time, allowing multiple tasks to run concurrently without monopolizing resources. Additionally, the scheduler manages context switching between tasks, saving and restoring their states as needed, while maintaining a job queue that optimizes the order of tasks based on priority and resource availability. It also distributes tasks across available processing units to optimize resource usage and prevent bottlenecks through load balancing, implements deadlock prevention strategies to avoid indefinite waiting for resources, and responds to events that may trigger the execution of specific tasks, such as user input or incoming messages.

*C. Modules*

Each module may be designed differently (Fig. 4) depending on the requirements: it could be a monolith, a set of services (as in SOA), a set of microservices, or a different design that may emerge in time. This means that architects are not required to choose only one of these options, but could design a multi-paradigm architecture.

As much has been written about monoliths, services, and microservices ($\mu$S), I won't delve into these topics in this paper.

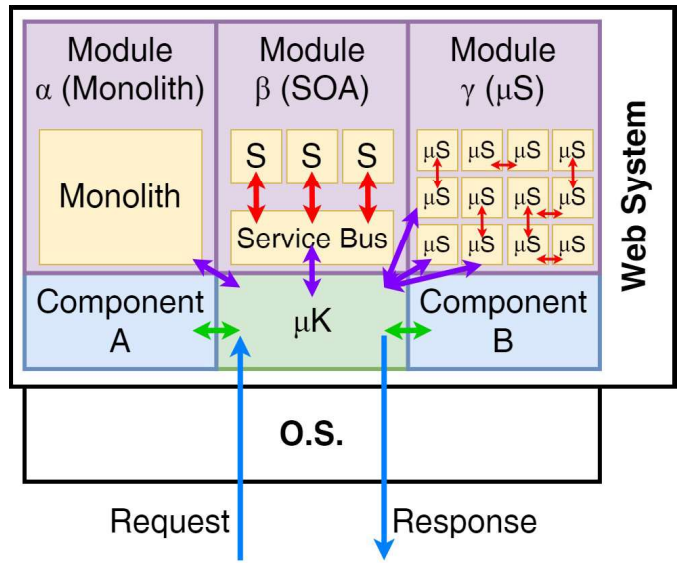

Fig. 4. Multi-Paradigm Modular Web System

## V. Self-* Properties

The $\mu$K architecture proposed in this paper must be inherently self-adaptive to support and manage multiple kinds of modules, and their respective implications. Evidently, we wouldn't want to hinder nor omit the advances made with orchestrators [20]. Therefore, it should offer the possibility to automatically scale in coherence with the system's user demand and its available resources, and to contemporaneously redirect, or balance the load of, user requests and responses. Although, in this case, the $\mu$K architecture will not only support containers, but also other execution environments, which should offer backward compatibility with older, purely monolithic systems.

Being this a more tightly coupled architecture than that used for microservices nowadays, it will be possible for modules to self-diagnose, self-update, and truly self-heal.

Currently, these self-* properties are being performed by orchestrators at the system level. More specifically, a container's *health* or responsiveness is periodically checked by the orchestrator via probes. Updates are performed with CI/CD pipelines whose result is an updated image that is later instantiated in containers, while the outdated containers are decommissioned or terminated. Similarly, when a container isn't responsive, it's restarted, or a new container is started while the faulty one is decommissioned or terminated, according to the policy in place.

Even though these strategies somehow allow the system to recover, the component (container/image) level is not actually diagnosing, updating, nor truly healing itself, as the underlying problems aren't really being solved.

For these reasons, I have applied the MAPE-K (Monitor, Analyze, Plan, Execute - Knowledge) loop [21] at the container/virtual machine (execution environment) level, as shown in Fig. 5. The novelties in this diagram include the fact that the monitor is aware of the inputs and outputs of the container/virtual machine (VM), in addition to its resource consumption. Also, a set of self-healing libraries is included in the container/VM to help it truly heal itself with the assistance of the planner and executor of the MAPE-K loop.

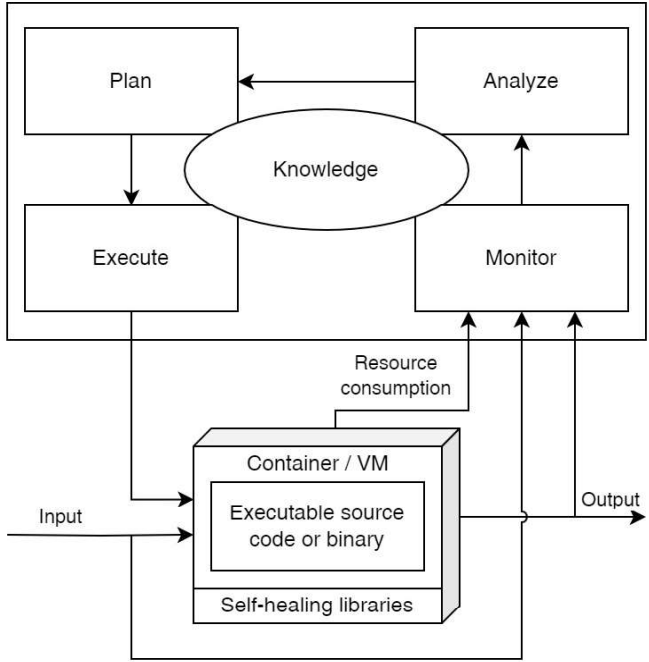

Fig. 5. MAPE-K Loop at the Container/VM Level

## VI. Tackling Microservice Challenges

Reconsidering the challenges listed in Section III, the $\mu$K architecture should help tackle them as follows:

1) **Security Risks**: The multi-paradigm features supported by the $\mu$K will allow designers to reapply historically proven solutions (i.e. from monoliths, SOA, and microservices) to their newer designs.
2) **Performance Overhead**: As will be further discussed in Section VII, $\mu$K-based web architectures are expected to have better performance than those solely based on microservices, but lower than purely monolithic ones. However, this remains to be demonstrated.
3) **Service Deployment and Monitoring**: Although deploying for multiple paradigms will certainly increase the complexity of this process, the fact that it will be managed by a $\mu$K shall make it more straightforward. Additionally, the reasons for manual intervention will decrease with the implementation of the aforementioned MAPE-K loop and the supporting logic.
4) **Service Interaction and Communication**: The Monitor and Service Manager of the $\mu$K will facilitate keeping up with new microservice additions to the system, even as it scales.
5) **Finding the Right Granularity**: The flexibility conveyed by this multi-paradigm architecture allows for experimentation to determine and select the best design solution, including a combination of paradigms if deemed convenient.
6) **Data Management**: Architects will be able to select centralized or distributed database systems and will have access into all of them, as they'll all be reachable via the $\mu$K.
7) **Troubleshooting**: Debugging will be facilitated by the $\mu$K and MAPE-K loop implementation (with the self-healing libraries), offering debuggers a greater reach throughout the web system and allowing them to locate faults much quicker.

## VII. Implementation Considerations

Having updated the design of the microkernel-based web architecture proposed in [4], one shall consider the technologies that will be used for its implementation. For example, [4] used PHP5. Nowadays, WebAssembly could be a more convenient solution, given its multilingualism and optimized virtual machine, which may be deployed on different kinds of (heterogeneous) devices, even with resource constraints [22], thus furthering the cloud-edge computing continuum. In particular, efforts are being made for the executables that derive from traditionally unsafe languages to become safe via solutions such as MSWASM [23].

Additionally, this architecture has been designed for the $\mu$K and its components to be in memory permanently, while its modules and their respective elements may be loaded into memory according to user demand.

## VIII. Conclusion & Future Work

Having identified the main challenges involved with microservice architectures, I proceeded to update the design of the microkernel-based web architecture proposed in [4] taking into account such challenges and the architectural developments since its writing. Afterwards, I analyzed current approaches to several self-* properties and how this novel architecture may improve them. Once the top-level design of the $\mu$K architecture was complete, I looked into how it may help tackle the aforementioned challenges. Finally, I reflected upon several implementation aspects of the work proposed in this paper.

As for the research questions, I provided an answer for RQ1 in Section III and for RQ2 throughout the rest of this document.

The next steps in this research line include:
- Implementing the microkernel and the components stipulated in this paper;
- Implementing a web system; and
- Benchmarking the web system that uses the microkernel.

## IX. Acknowledgement

This work is partially supported by the Italian Ministry of University and Research under the PNRR program (financed by the EU, NextGenerationEU), Ministerial Decree n. 352, and Atlante SRL.